\documentclass[aps,preprint]{revtex4}%
\usepackage{amsfonts}
\usepackage{amsmath}
\usepackage{amssymb}
\usepackage{graphicx}%
\begin{document}
\preprint{ }
\title[Tunneling failure at low frequencies]{The tunneling model of laser-induced ionization and its failure at low frequencies}
\author{H. R. Reiss}
\affiliation{Max Born Institute, 12489 Berlin, Germany}
\affiliation{Physics Department, American University, Washington, DC 20016-8058, USA}


\begin{abstract}
The tunneling model of ionization applies only to longitudinal fields:
quasistatic electric fields that do not propagate. Laser fields are
transverse: plane wave fields that possess the ability to propagate. Although
there is an approximate connection between the effects of longitudinal and
transverse fields in a useful range of frequencies, that equivalence fails
completely at very low frequencies. Insight into this breakdown is given by an
examination of radiation pressure, which is a unique transverse-field effect
whose relative importance increases rapidly as the frequency declines.
Radiation pressure can be ascribed to photon momentum, which does not exist
for longitudinal fields. Two major consequences are that the near-universal
acceptance of a static electric field as the zero frequency limit of a laser
field is not correct; and that the numerical solution of the
dipole-approximate Schr\"{o}dinger equation for laser effects is inapplicable
as the frequency declines. These problems occur because the magnetic component
of the laser field is very important at low frequencies, and hence the dipole
approximation is not valid. Some experiments already exist that demonstrate
the failure of tunneling concepts at low frequencies.

\end{abstract}
\date[7 May 2014]{}
\email{reiss@american.edu}
\maketitle

\section{Introduction}

The quantum phenomenon of tunneling through a potential barrier has been known
and fruitfully employed since its introduction in 1928 to describe nuclear
alpha decay\cite{gamow} and, in the same year, to calculate the ionization of
the hydrogen atom by a constant electric field\cite{oppie}. Both of those
early applications involved static electric fields. The same concept has been
applied in more recent years\cite{keldysh,n+r,ppt} to laser-induced ionization
when the laser field is approximated as a quasistatic electric field. The
concept of tunneling ionization is that an impenetrable potential barrier is
rendered penetrable by the superposition of an oscillatory electric field:
$\mathbf{E}\left(  t\right)  $. It is usual to describe the applied field by a
scalar potential,
\begin{equation}
\phi\left(  t\right)  =-\mathbf{r\cdot E}\left(  t\right)  ,\;\mathbf{A}=0,
\label{a}%
\end{equation}
since this leads to the familiar graphical illustration where a potential well
representing the force binding an electron to an atom is periodically
depressed to allow the electron to escape by tunneling through the depressed
barrier. The potentials in Eq.(\ref{a}) describe a \textit{longitudinal
field}. Longitudinal fields can oscillate in time, but they do not propagate.
The scalar-potential-only nature of Eq.(\ref{a}) is usual but not essential.
It is always possible to find a gauge transformation that replaces the scalar
potential $\phi\left(  t\right)  $ in Eq.(\ref{a}) by a vector potential
$\mathbf{A}\left(  t\right)  $. Gauge equivalence means that such a
vector-potential description also represents a longitudinal field.

Actual laser fields are true propagating plane-wave (PW) fields, also known as
\textit{transverse fields}. The name comes from the fact that a PW field has
electric and magnetic components of equal magnitude (in Gaussian units) that
are perpendicular to each other, and the plane they define is perpendicular to
the direction of propagation. That is, propagation is in a direction
transverse to the electric and magnetic fields. Laser fields are
\textit{vector fields} that cannot be completely described by a scalar
potential, nor by any potential that is gauge-equivalent to a scalar
potential. The simplest way to describe transverse fields (see, for example,
the textbook of Jackson\cite{jackson}) is by a vector potential function
alone
\begin{equation}
\phi=0,\;\mathbf{A}=\mathbf{A}\left(  \varphi\right)  \label{a1}%
\end{equation}
that depends on spacetime coordinates $x^{\mu}$ only in the combination%
\begin{align}
\varphi &  =k^{\mu}x_{\mu}=\omega t-\mathbf{k\cdot r,}\label{b}\\
k^{\mu}  &  :\left(  \frac{\omega}{c},\mathbf{k}\right)  ,\quad x^{\mu
}:\left(  ct,\mathbf{r}\right)  . \label{b1}%
\end{align}
The quantity $\varphi$ is the phase of a propagating field, and $k^{\mu}$ and
$\mathbf{k}$ are the 4-vector propagation vector and its 3-vector component.
This prescription for the potentials is known as the \textit{radiation gauge}
(or \textit{Coulomb gauge}).

Longitudinal and transverse fields are fundamentally different electromagnetic
phenomena. These differences are explored in depth in this article.

For some purposes it is possible to neglect the dependence of $\mathbf{A}$ on
the spatial coordinate $\mathbf{r}$, in which case (known as the
\textit{dipole approximation}) there is a gauge transformation due to
G\"{o}ppert-Mayer\cite{gm} (GM) that provides a gauge equivalence between the
dipole-approximation vector potential $\mathbf{A}\left(  t\right)  $ and the
scalar potential of Eq.(\ref{a}). It is of fundamental importance to maintain
awareness that fields described within the GM gauge (also known as the
\textit{length gauge}) cannot be anything more than longitudinal fields. A
longitudinal field can never be gauge-equivalent to a transverse field, but
only to an approximation to a transverse field. This means that the tunneling
concept, dependent on the scalar potential (\ref{a}) or any potential
gauge-equivalent to (\ref{a}), can never be more than a limited approximation
for laser phenomena. It is the elucidation of these limitations that is a
major focus of this article.

Tunneling is a concept applicable only to longitudinal fields, and is only a
very limited approximation for transverse (e.g. laser) fields. The focus of
attention is now shifted to an examination of parameters wherein tunneling can
be a meaningful approximation for laser fields.

There is an upper limit on the field frequency for which the dipole
approximation is applicable that was pointed out by G\"{o}ppert-Mayer\cite{gm}%
. When the field wavelength is less than the size of the atom, it can act as a
probe of the structure of the atom. That is not possible for $\lambda
\gtrsim1\;a.u.$, which limits the field frequency to
\begin{equation}
\omega\lesssim2\pi c \label{c}%
\end{equation}
in atomic units.

It is not generally recognized that there is a lower limit to the frequency at
which the dipole approximation can be applied. This lower limit is of far more
practical importance than the upper limit of Eq.(\ref{c }). The reason for
this oversight may be that no such lower limit exists for QSE fields but,
importantly, it does for PW fields

Section II gives a qualitative insight into the atomic domain as it appears
for QSE fields; that is for fields describable by Eq.(\ref{a}), and Section
III repeats the analysis for PW fields. The contrast between the two types of
fields is striking. This comes about because the strength of a longitudinal
field is judged only by the magnitude of the electric field, whereas it is the
strongly frequency-dependent ponderomotive potential that is required for
appraisal of transverse fields\cite{hr62b,hr89}. For instance, a parameter
domain that applies to very weak QSE fields is shown to consist of very strong
PW fields, and vice versa. Important matters elaborated in Section III include
the criteria for the onset of nondipole effects at low frequencies. Section IV
concentrates explicitly on low-frequency behavior.

The analytical methods available for the description of ionization by strong
laser fields are described in Section V. These include the tunneling model,
the strong-field approximation (SFA), and the numerical solution of the
time-dependent Schr\"{o}dinger equation (TDSE). The domains of applicability
have some overlap, but this overlap is far more limited than is evident from
the current literature. In particular, the literature exhibits widespread
dependence on the \textquotedblleft exactness\textquotedblright\ of the TDSE
without regard for the failure of the dipole approximation for low-frequency
laser fields.

Finally, the results are summarized and evaluated in Section VI. Qualitative
assessments are made, including the conclusion that the tunneling
approximation is useful for laser-induced ionization problems in only a very
small region in the frequency and intensity domain of laser fields.

\section{Quasistatic electric fields}

The simplest electromagnetic field is a static electric field. This can be
described by the potentials%
\begin{equation}
\phi=-\mathbf{r\cdot E}_{0},\;\mathbf{A}=0, \label{d}%
\end{equation}
where the subscript on the electric field vector $\mathbf{E}_{0}$ is a
reminder that the field so described is constant. A simple but important
generalization is a field configuration in which there is no magnetic
field\ and the electric field is time-dependent, as in Eq.(\ref{a}). This a
quasistatic electric (QSE) field. Another indicator of this identification is
the Lorentz invariant%
\begin{equation}
\mathbf{E}^{2}-\mathbf{B}^{2}=-\frac{1}{2}F^{\mu\nu}F_{\mu\nu}, \label{e}%
\end{equation}
where the inner product of the two electromagnetic-field tensors $F^{\mu\nu}$
on the right-hand side of the equation shows the reason why this quantity is a
Lorentz scalar; that is, its value is invariant under any Lorentz
transformation. For QSE fields, it is always true that%
\begin{equation}
\mathbf{E}^{2}-\mathbf{B}^{2}>0. \label{f}%
\end{equation}
By contrast, a laser field is a PW field for which it is always true that%
\begin{equation}
\mathbf{E}^{2}-\mathbf{B}^{2}=0. \label{g}%
\end{equation}
The important conclusion is that the GM gauge, that employs the potentials
(\ref{a}) (or any potentials gauge-equivalent to (\ref{a})) for the
description of a laser field, approximates the laser field by a QSE field.
There is no limit in which the GM gauge is exact.%

\begin{figure}
[ptb]
\begin{center}
\includegraphics[
trim=0.208537in 0.000000in 0.000000in 0.000000in,
height=5.3843in,
width=6.7914in
]%
{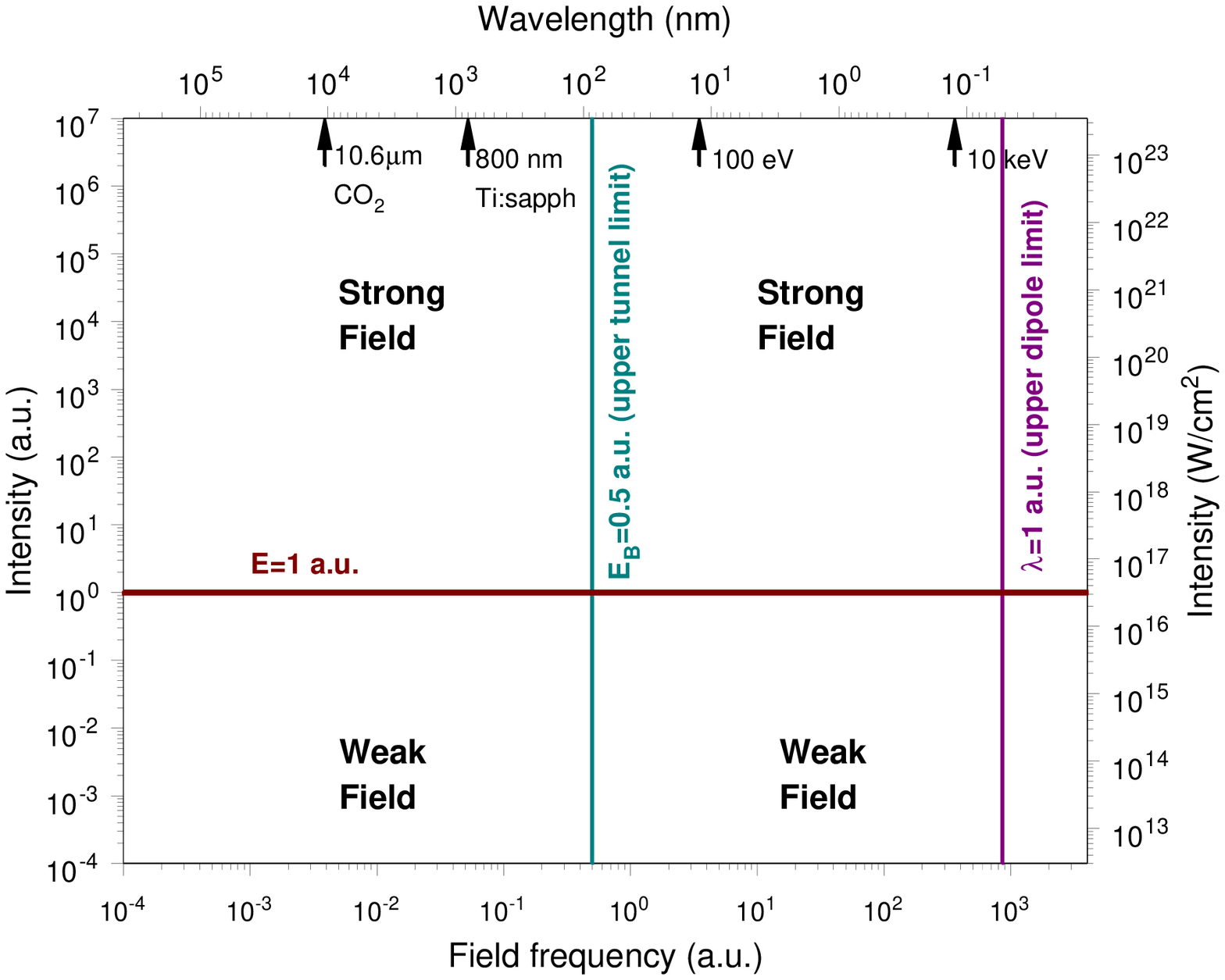}%
\caption{A QSE field contains field information solely in the form of the
direction and amplitude of the electric field $\mathbf{E}$. For atomic
problems, $\left\vert \mathbf{E}\right\vert \gg1a.u.$ indicates a strong field
and $\left\vert \mathbf{E}\right\vert \ll1a.u.$ indicates a weak field. No
other field information appears in the figure. The location of the atomic
binding energy $E_{B}$ (selected to be $0.5a.u.$) is shown, since a tunneling
model applies only if the field frequency is such that $\hbar\omega\ll E_{B}$.
The location of the frequency corresponding to wavelength $\lambda=1a.u.$ is
also shown, since it is known that the dipole approximation will not be valid
beyond that frequency.}%
\label{fg1}%
\end{center}
\end{figure}

A\ QSE field is a longitudinal field, which means that there is only one
spatial direction that is important: the electric field direction. A\ QSE
field does not possess a propagation capability; it can oscillate with time,
but it cannot propagate. The electromagnetic field enters into the potentials
through the direction and the magnitude of the electric field. In atomic
applications, the electric field is strong when $\left\vert \mathbf{E}%
\right\vert \gg1a.u.$, and it is weak when $\left\vert \mathbf{E}\right\vert
\ll1a.u.$ This is indicated in Fig. \ref{fg1}, which is a plot showing field
frequency on the $x-axis$ and field intensity on the $y-axis$. Intrinsic
frequency considerations do not occur, but two frequency limits of atomic
origin are shown. One is the well-known upper limit on the dipole
approximation, where Fig.\ref{fg1} shows the frequency that is given in
Eq.(\ref{c}). Another upper limitation on the frequency comes from the
tunneling method itself, since tunneling is only meaningful when an
uncountably large number of photons participate. In practical terms,
\textquotedblleft uncountably large\textquotedblright\ in an ionization event
may be satisfied by a number of the order of\ $10$. Figure \ref{fg1} shows
this tunneling limit in terms of the binding energy of an electron in
ground-state hydrogen at $0.5a.u.$

Among the qualitative properties just listed about QSE fields, there is
nothing to serve as an indicator that the dipole approximation is inapplicable
to laser fields at low frequencies. This is apparently the underlying reason
why a low-frequency limit has escaped notice in the Atomic, Molecular, and
Optical (AMO) community. Even a modern book entitled \textit{Atoms in Intense
Laser Fields}\cite{jkp} asserts (see pp.267-289) that the GM gauge is the
preferred gauge for laser problems since it is well-behaved as the frequency
approaches zero. There is no awareness that there is a low-frequency limit for
the applicability of the GM gauge to laser problems. The line of reasoning
followed by the authors is based on the concept of adiabaticity, and the
entire discussion depends on the validity of the dipole approximation. The
fact that a constant electric field emerges as $\omega\rightarrow0$ is
\textit{prima facie} evidence that the discussion in Ref.\cite{jkp} relates
only to longitudinal fields and not to laser fields.

Another clear indication of the problem that exists in the AMO community can
be found in a recent paper in a prestigious rapid-publication journal, where
two consecutive sentences that are contradictory are viewed as if they were
mutually supportive\cite{arissian}. The first sentence is: \textquotedblleft
In adiabatic tunneling the laser field is treated as if it were a static
field, time serving only as a parameter.\textquotedblright\ That is, the
authors state that they are treating the laser field as if it were a QSE
field, where Eq.(\ref{f}) is valid. The next sentence is: \textquotedblleft It
is rigorously valid for long wavelengths ...\textquotedblright\ The authors
thereby state that they view the laser field as if it has a zero-frequency
limit, and they view that limit as a static electric field, despite the
previous statement that they are concerned with laser fields, where
Eq.(\ref{g}) applies. Equations (\ref{f}) and (\ref{g}) are incompatible, not
equivalent. References \cite{jkp} and \cite{arissian} are not singled out for
special criticism, but they are cited because they are especially visible
representations of the prevailing view.

The following Section further elaborates the fundamental differences between
QSE fields and laser fields.

\section{Plane wave fields}

Plane wave (PW) fields are transverse fields. (See Chapter 7 in the text by
Jackson\cite{jackson}.) An essential property of PW fields is that any
occurrence of the spacetime 4-vector $x^{\mu}$ must occur as the scalar
product with the propagation 4-vector $k^{\mu}$\cite{schwinger,s+s,hrjmo}, as
shown in Eq.(\ref{b}). This product, which is the phase of a propagating
field, will be called $\varphi$. The descriptions \textquotedblleft
PW\textquotedblright, \textquotedblleft transverse\textquotedblright, and
\textquotedblleft propagating\textquotedblright\ are here viewed as equivalent
designations of the type of fields that lasers produce. A basic property of
such fields is that, once generated, they propagate indefinitely in vacuum
without the need for sources.

This last property is very important. The GM gauge does not have that feature
of a freedom from sources. That is, no field can exist in the GM gauge without
sources to sustain it\cite{hrjpb13}. They are \textquotedblleft virtual
sources\textquotedblright\ in the sense that they do not actually exist in the
laboratory. Because of the gauge equivalence between the GM gauge and a dipole
approximation to a PW, the properties of those sources do not normally intrude
in a calculation. There are special cases, however, when the virtual
sources\ of the GM gauge can produce unintended consequences\cite{hrjpb13}
even when the dipole approximation is valid.

The best indicator of the strength of coupling of a PW field to a charged
particle is the ponderomotive potential $U_{p},$ given by%
\begin{equation}
U_{p}=I/\left(  2\omega\right)  ^{2} \label{h}%
\end{equation}
in atomic units, where $I$ is the field intensity. In the radiation gauge,
this quantity is a true potential energy\cite{hr89} that depends on the local
values of $I$ and $\omega$. Its dimensionless form also serves as the coupling
constant between a strong PW field and an electron, replacing the fine
structure constant $\alpha$ of perturbation theory\cite{hr62a,hr62b,hr89}. The
analog \ of $U_{p}$ in the GM gauge is a \textquotedblleft quiver
energy\textquotedblright\ that corresponds to the oscillation of the charged
particle as it is driven by the QSE field. The magnitude of $U_{p}$ is the
same in both gauges, but the interpretation is different. In the GM gauge,
$U_{p}$ is a kinetic energy of the charged particle that results from the
virtual sources described above. In the radiation gauge, there are no sources
to drive the particle, and $U_{p}$ is a potential energy, not a kinetic energy.%

\begin{figure}
[ptb]
\begin{center}
\includegraphics[
trim=0.222440in 0.000000in 0.000000in 0.000000in,
height=5.3843in,
width=6.7784in
]%
{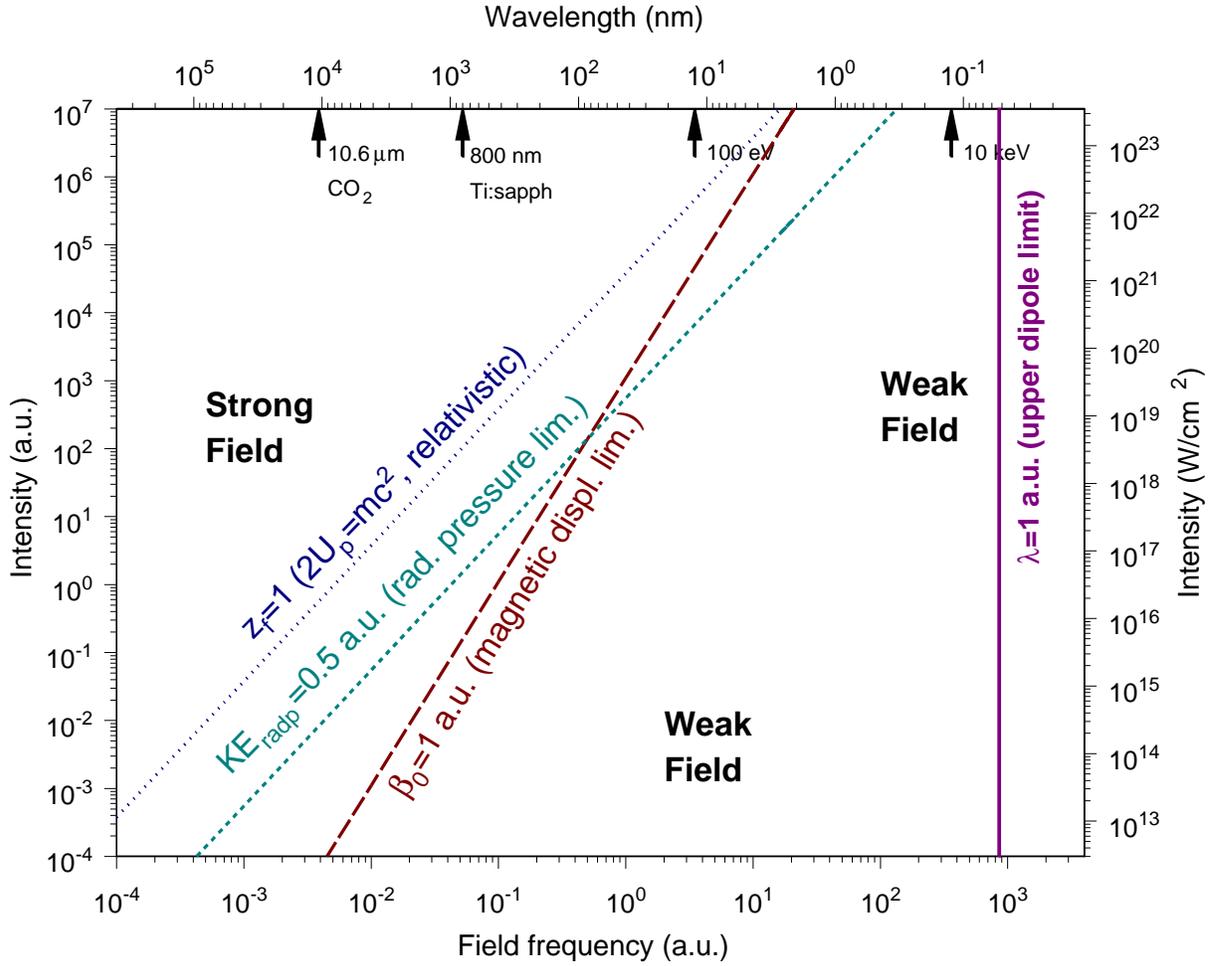}%
\caption{A\ PW field is a propagating field that has many more physical
features than a QSE field. The most important properties to note are the major
qualitative differences between Fig.\ref{fg1} and Fig.\ref{fg2}. Relativistic
conditions are clearly present here, including regions such as in the lower
left of the figure where relativistically-strong-field effects occur in
regions that are labeled \textquotedblleft Weak Field\textquotedblright\ in
Fig.\ref{fg1}. The \textquotedblleft Weak Field\textquotedblright\ designation
in the upper right of this figure contrasts with the \textquotedblleft Strong
Field\textquotedblright\ label in Fig.\ref{fg1}. The onset of low-frequency
nondipole behavior is marked either by the displacement in the field
propagation direction caused by the magnetic field ($\beta_{0}=1a.u.$) or by
radiation-pressure-caused contributions to energy and momentum ($KE_{radp}%
=0.5a.u.$). Neither of these low-frequency indicators exists in Fig.\ref{fg1}.
The overall conclusion is that QSE fields and PW fields are fundamentally
different electromagnetic phenomena with major qualitative differences.}%
\label{fg2}%
\end{center}
\end{figure}

In the nonperturbative theory of the interaction of charged free particles
with strong laser fields, as in Compton scattering\cite{sengupta},
photon-multiphoton pair production\cite{hr62a}, or pair
annihilation\cite{n+r64}, a single intensity parameter occurs that is the same
for all free-particle interactions. Many different notations occur in the
literature, but it will be designated here as $z_{f}$, where%
\begin{equation}
z_{f}=2U_{p}/mc^{2}. \label{i}%
\end{equation}
The quantity $z_{f}$ can be viewed as a dimensionless statement of the
ponderomotive potential. When $z_{f}=1$, then $2U_{p}$ equals the rest energy
of the particle, and the process is unequivocally relativistic. This means
that the dipole approximation has no validity. Figure \ref{fg2} is the PW
analog of Fig.\ref{fg1} for QSE fields, and the line corresponding to
$z_{f}=1$ shows immediately that there is a drastic difference between the
behavior of QSE and PW fields. For example, the lower left region of
Fig.\ref{fg1} corresponds to a very weak QSE field, whereas the same region in
Fig.\ref{fg2} represents a very strong PW field. The line corresponding to
$z_{f}=1$ is given by
\begin{equation}
I_{rel}=2c^{2}\omega^{2} \label{i0}%
\end{equation}
in atomic units.

\begin{figure}
[ptb]
\begin{center}
\includegraphics[
trim=0.000000in 1.881585in 0.000000in 1.255382in,
height=2.2805in,
width=6.2076in
]%
{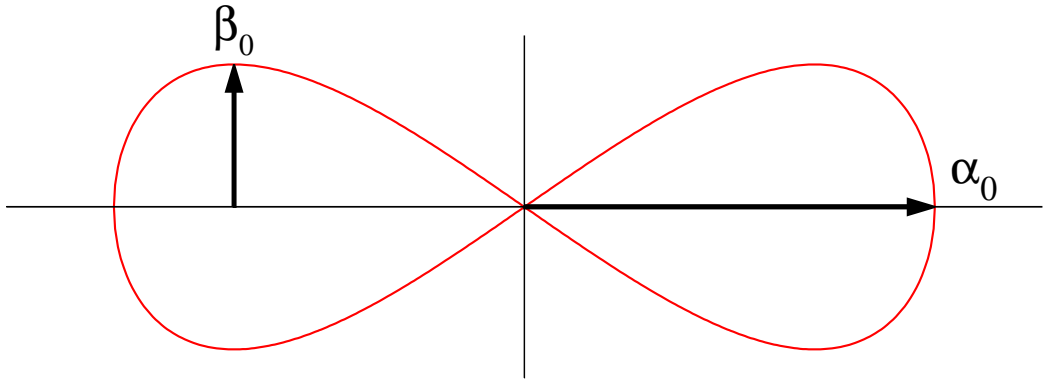}%
\caption{A free electron in a PW field executes a \textquotedblleft
figure-8\textquotedblright\ orbit in a frame of reference where the electron
is at rest on average. The amplitude $\alpha_{0}$ is in the direction of the
electric field. The amplitude of motion in the direction of propagation of the
PW\cite{l+l,s+s}, $\beta_{0}$, caused by the combined action of the electric
and magnetic fields, is defined as shown here. When $\beta_{0}$ is of the
order of one $a.u.$, this is an indication of the low frequency failure of the
dipole approximation.}%
\label{fg3}%
\end{center}
\end{figure}

The condition $z_{f}=1$ refers to a strongly relativistic environment, but the
onset of nondipole behavior can occur at significantly lower intensities. One
way to estimate the lower-frequency limit of the dipole approximation is to
examine the well-known \textquotedblleft figure-8\textquotedblright\ motion of
a free charged particle in a PW field\cite{l+l,s+s}. This is shown in
Fig.\ref{fg3} in the frame of reference where the particle is at rest when
averaged over a full cycle. At low field intensity, the figure-8 reduces to a
straight-line oscillation of amplitude $\alpha_{0}$. Departure from
straight-line behavior occurs at increasing intensity since the coupled action
of the electric and magnetic fields of the PW causes a motion in the direction
of propagation of the field. When the amplitude in the propagation direction
is of the order of one atomic unit,
\begin{equation}
\beta_{0}\approx\frac{U_{p}}{2mc\omega}=1a.u., \label{i1}%
\end{equation}
this signals an important contribution from the magnetic field that will
certainly influence the nature of the interaction with the ion, and the dipole
approximation is not valid. The line determined by the condition (\ref{i1})
is
\begin{equation}
I_{fig8}=8c\omega^{3} \label{i2}%
\end{equation}
in atomic units, and is shown in Fig.\ref{fg2}.%

An alternative way to assess the importance of nondipole effects is to
consider the momentum and energy of a motion induced by radiation pressure.
The momentum in the direction of field propagation resulting from radiation
pressure is\cite{hrrel,t+d,hr87}%
\begin{equation}
p_{\shortparallel}=U_{p}/c. \label{j}%
\end{equation}
The kinetic energy corresponding to this momentum is%
\begin{equation}
KE_{radp}=\frac{p_{\shortparallel}^{2}}{2m}=\frac{U_{p}^{2}}{2mc^{2}},
\label{k}%
\end{equation}
where the nonrelativistic form on the left-hand side is sufficient since this
limitation occurs well before the fully relativistic limit is reached at
$2U_{p}=mc^{2}.$ The condition (\ref{k}) gives the line in Fig.\ref{fg2}
determined by%
\begin{equation}
I_{radp}=4c\omega^{2} \label{l}%
\end{equation}
in atomic units. This is parallel to the $I_{rel}$ of Eq.(\ref{i0}), but
smaller by the factor $2/c$ a.u. That is, the onset of a relativistic effect
like radiation pressure makes its presence felt well before the fully
relativistic condition of Eq.(\ref{i0}).

By using very sensitive measurement techniques, radiation pressure effects
have already been observed in the laboratory\cite{smeenk} at 800 nm
wavelength. An intensity of $8\times10^{14}W/cm^{2}$ was specifically
analyzed\cite{t+d,hr87}. In atomic units, the intensity was $0.023$, whereas
Eq.(\ref{l}) predicts $I_{radp}=1.78a.u.$ for that wavelength. Thus the
experiment detected radiation pressure at little more than one percent of the
condition stated in Eq.(\ref{l}). This confirms that Eq.(\ref{l}) gives a
realistic assessment of conditions where the dipole approximation will fail.

\section{Low Frequencies}

Figures \ref{fg1} and \ref{fg2} make it very clear that longitudinal fields
and transverse fields are different electromagnetic phenomena. That major
distinction arises from the unique properties of a propagating field. The
failure of correspondence is most striking at low frequencies. From the point
of view of longitudinal fields, low frequencies are viewed as being in the
so-called \textit{tunneling limit}, where QSE fields approach static electric
fields. For transverse fields, low frequencies correspond to the completely
different domain of Extreme Low Frequency (ELF) radio waves. The qualitative
features of these two fundamentally different domains of electromagnetic
phenomena are outlined here.

\subsection{\textquotedblleft Tunneling limit\textquotedblright\ and the
Keldysh parameter}

The tunneling view of ionization leads to a single controlling intensity
parameter known as the Keldysh parameter\cite{keldysh} that can be written as%
\begin{equation}
\gamma_{K}=\sqrt{E_{B}/2U_{p}}, \label{l0}%
\end{equation}
where $E_{B}$ is the binding energy of the electron in the atom or molecule.
This quantity is also called the ionization potential, and designated by $IP$
or $I_{p}$. The putative \textit{tunneling domain} is defined by%
\begin{equation}
\gamma_{K}<1, \label{l1}%
\end{equation}
and the \textit{tunneling limit} is%
\begin{equation}
\gamma_{K}\rightarrow0. \label{l2}%
\end{equation}
The opposite case to the tunneling domain is the so-called \textit{multiphoton
domain}: $\gamma_{K}>1$. It has become standard practice in the description of
laser-induced processes to specify whether an experiment or theory relates to
one or the other of these two longitudinal-field domains, without regard to
the qualitatively different behavior of actual laser (transverse) fields.

With Eq.(\ref{h}) used to introduce field intensity and frequency into
Eq.(\ref{l0}), the dividing line between these two domains will be called
$I_{tun}$, and is given by%
\begin{equation}
I_{tun}=2E_{B}\omega^{2}, \label{l3}%
\end{equation}
which leads to the tunneling limit expressed as $I_{tun}\gg2E_{B}\omega^{2}$.
Equation (\ref{l3}) has the same $\omega^{2}$ dependence as $I_{rel}$ in
Eq.(\ref{i0}) and $I_{radp}$ in Eq.(\ref{l}), although the prefactors are such
that $I_{tun}=1$ lies well below the other two lines in a diagram such as
Fig.\ref{fg2}. As the frequency declines, however, the physical system heads
inexorably towards the relativistic domain where the dipole approximation
certainly fails and the tunneling approximation has no applicability to laser
effects\cite{hr82}. On the other hand, from the point of view of longitudinal
fields, declining frequency is simply a matter of an ac field approaching a dc
limit. This appears to present no conceptual problems\cite{shakeshaft}, and
the AMO community routinely evaluates analytical methods by the limit as
$\omega\rightarrow0$. If that limit corresponds to static-electric-field
results, this is regarded as evidence of an acceptable theory. The fact that
$\mathbf{E}^{2}-\mathbf{B}^{2}>0$ for such a limit seems never to be noticed.

\subsection{ELF radio waves}

Transverse fields always retain their propagation property as $\omega
\rightarrow0,$ meaning that $\mathbf{E}^{2}-\mathbf{B}^{2}=0$ is sustained and
the magnetic field is always present. Maintenance of the propagation property
means that decreasing frequency implies a progression from ultraviolet to
visible to infrared to microwave to radio phenomena. The zero frequency limit
can be approached, but never reached. The divergence of $U_{p}$ as
$\omega\rightarrow0$, shown in Eq.(\ref{h}), is symptomatic of the energy
demands of producing ELF radio transmissions. The lowest ELF frequency of
which this writer is aware is the $76$ $Hz$ system that the U.S. Navy proposed
as a means of communicating with submerged submarines. The project was named
\textquotedblleft Project Sanguine\textquotedblright\cite{sanguine}, and the
original design (never built) required a massive $600$ $MW$ of power to
produce a signal with such small bandwidth that only simple coded messages
could be sent.

An ELF radio wave is conceptually as distant from a constant electric field as
would be a pure magnetic field when judged by the relevant $\mathbf{E}%
^{2}-\mathbf{B}^{2}$ value. Nevertheless, the hallmark of success valued from
a QSE point of view in the AMO community is that theoretical predictions
should match the properties of constant electric fields. This is completely
inappropriate for laser fields.

\section{Analytical methods}

A brief survey is given here of some of the consequences of the foregoing
considerations for theoretical techniques employed to describe atomic systems
subjected to very intense laser fields.

\subsection{Tunneling}

Time-independent potential barriers that are penetrable by quantum tunneling
processes are treated in essentially all textbooks on quantum mechanics, and
that problem is thoroughly understood. Tunneling methods were later applied to
QSE fields\cite{keldysh,n+r,ppt}. Many investigators who depend on tunneling
methods to solve problems in laser-induced ionization processes regard the
existence of a static-field limit of their method as a reassurance of
accuracy. This should be worrisome rather than reassuring. Transverse fields
do not have a physically attainable zero frequency limit, which is evident in
several ways. The radiation pressure result would be divergent were there a
zero frequency limit, as is clear from Eqs.(\ref{j}) and (\ref{h}). Figure
\ref{fg2} shows that there is no access to zero frequency of a PW field
without entering into the relativistic domain where the dipole approximation
necessarily fails. Under relativistic conditions the magnetic component of a
transverse field becomes as important as the electric component. Theories of
relativistic tunneling have been published\cite{kmp}, but they relate only to
extremely strong QSE fields. Relativistically strong laser fields cannot be
treated by QSE methods.%

\begin{figure}
[ptb]
\begin{center}
\includegraphics[
height=4.523in,
width=5.8608in
]%
{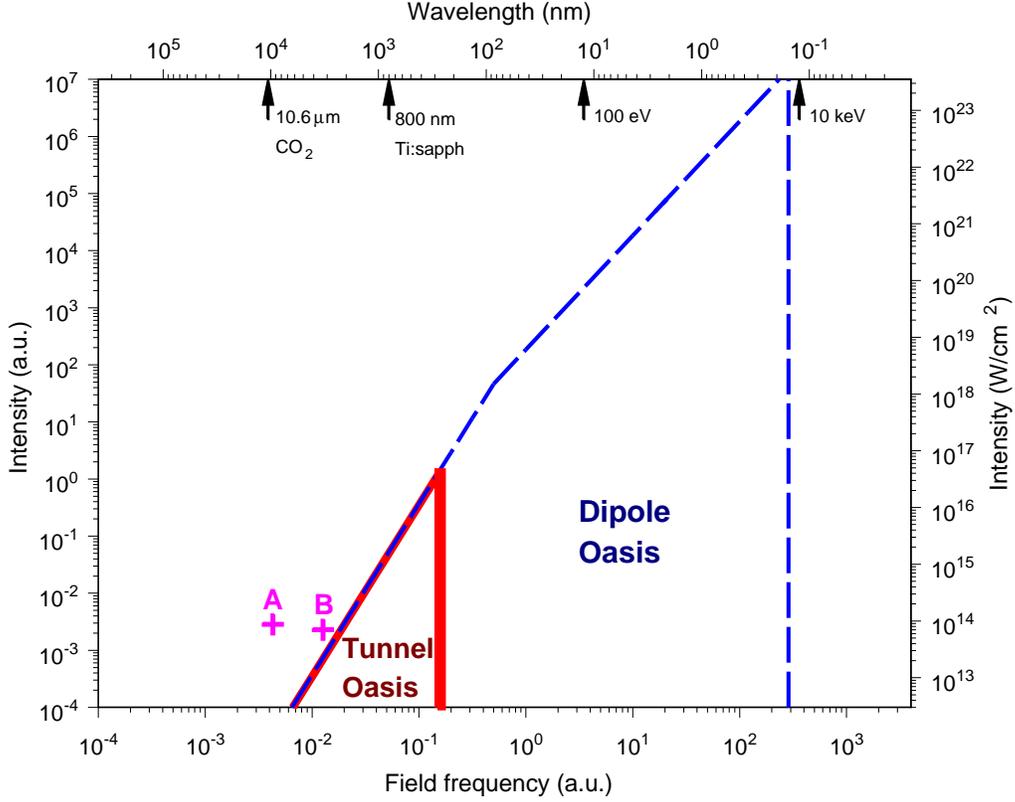}%
\caption{This figure shows the domain of applicability of the tunneling model
for laser-induced ionization -- dubbed \textquotedblleft Tunnel
Oasis\textquotedblright\ -- that follows from applying the constraints shown
in Fig.\ref{fg2}. The Tunnel Oasis is enclosed in the triangular domain
bounded by solid lines. The Tunnel Oasis is a subset of the substantially
larger \textquotedblleft Dipole Oasis\textquotedblright, enclosed within the
dashed lines in the figure. The Dipole Oasis is the domain in which the dipole
approximation is applicable. The bounds at low frequencies all arise from the
constraints evaluated by examining radiation pressure effects in terms of
displacements due to the magnetic component of the laser field ($\beta
_{0}=1a.u.$) in Fig.\ref{fg2} or from the energy directly due to radiation
pressure effects ($KE_{radp}=0.5a.u.$) shown in Fig.\ref{fg2}. The
Dipole\ Oasis also represents that domain of field parameters in which there
is a gauge correspondence between the length and velocity gauges. Both Oases
expand considerably at lower intensities -- the domain of traditional AMO
physics. The point labeled \textquotedblleft$A$\textquotedblright\ locates the
experiments done with a $CO_{2}$ laser\cite{laval}, and the point
\textquotedblleft$B$\textquotedblright\ marks some of the
experiments\cite{osu1} done by the Ohio State University group. Both points
lie outside the Tunnel Oasis, and both sets of experiments show features that
are not explicable in terms of a tunneling process.}%
\label{fg4}%
\end{center}
\end{figure}

Figure \ref{fg4} shows the domain in an intensity-frequency diagram where the
tunneling method is applicable for laser-induced processes. This domain is
dubbed the \textquotedblleft Tunnel Oasis\textquotedblright, since that is
where a tunneling model can be applied successfully without concern for the
broader limitations of the tunneling approximation. The lower limits on
frequency come entirely from transverse-field effects as shown in
Fig.\ref{fg2}. Such effects are not present in a longitudinal-field analysis,
as Fig.\ref{fg1} clearly shows. The upper limit on frequency is specific to
tunneling, since the basic premise of a tunneling theory is that it is a
field-emission effect, not explicable in terms of a limited number of photons.

Experiments done with a $CO_{2}$ laser in the 1980s\cite{laval} showed a
striking departure from the properties of tunneling behavior\cite{hr102}. The
$CO_{2}$-laser parameters are shown as the point labeled \textquotedblleft%
$A$\textquotedblright\ in Fig.\ref{fg4}. This is plainly outside the Tunnel
Oasis. The pioneering \textquotedblleft low-energy-structure\textquotedblright%
\ (LES) experiments\cite{osu1,osu2} of the group at Ohio State University also
possess spectrum features that are not explicable within the tunneling
approximation, which is to be expected from the location marked
\textquotedblleft$B$\textquotedblright\ in Fig.\ref{fg4}. Both of these sets
of experiments with linearly polarized light show clear departures from
tunneling behavior, where the spectrum peaks sharply at zero
energy\cite{cbb,dk}.

A remarkable feature of the Tunnel Oasis is how small it is when compared to
the overall parameter space which is, or will become, the range of laser
parameters. One may call it an \textquotedblleft accident of
Nature\textquotedblright\ that the most commonly used source of strong laser
fields operates at about $800nm$, within the Oasis except at extremely large
intensities where saturation will occur, and the failure of the tunneling
model is obscured.

Another aspect of Fig.\ref{fg4} is that the lowest intensity shown is
$10^{-4}a.u.$ ($3.5\times10^{12}W/cm^{2}$). Traditional AMO physics is
conducted at much lower intensities than that, where the triangular Tunnel
Oasis expands considerably. This is the underlying reason why the limitations
of the GM gauge and its low-frequency failure have not become visible until recently.

\subsection{Strong-Field Approximation}

The Strong-Field Approximation (SFA) is based on the idea that after a
photoelectron has been ionized from an atom by a very intense field, its
behavior is dominated by the field that caused the ionization, rather than by
the residual effects of the binding potential\cite{hr80}. Discussion of the
SFA is difficult because its definition has become muddled. The analytical
approximation method of Ref.\cite{hr80} is not a tunneling method, so the name
\textquotedblleft Strong-Field Approximation\textquotedblright\ was
proposed\cite{hr90} for this method to distinguish it from GM-gauge methods
that are all basically tunneling approximations. Unfortunately, the purpose of
Ref.\cite{hr90} was not understood, and so the designation \textquotedblleft
SFA\textquotedblright\ began to be applied to all approximations where the
field in the final state dominates residual Coulomb effects.

The remarks that follow pertain only to the non-tunneling SFA of
1980\cite{hr80}. The 1990 paper\cite{hr90} showed that the method of
Ref.\cite{hr80} is actually based on a completely relativistic formalism, when
that can be subjected to the dipole approximation. The distinction that is
vital is that the magnetic field is always present but, apart from the
essential propagation property it imparts, its direct effect can otherwise be
ignored when nonrelativistic conditions obtain. This confines the 1980 SFA to
the portion of Fig.\ref{fg4} labeled \textquotedblleft Dipole
Oasis\textquotedblright, but not to the much smaller Tunnel Oasis. (Actually,
the SFA appears to retain some of its relativistic character even into the
low-frequency domain below the Tunnel Oasis\cite{hr102}.) This equivalence to
a full transverse-field approximation was demonstrated explicitly in another
paper, also in 1990\cite{hrrel}, by a completely relativistic calculation that
reduced in the nonrelativistic case to the 1980 SFA directly with no resort to
any tunneling-type approximations. The applicability of the 1980 SFA to
domains outside the Tunnel Oasis, but within the Dipole Oasis has been
verified by successful high frequency comparisons with the TDSE\cite{bondar}
and with the High-Frequency Approximation (HFA) of Gavrila\cite{hrjosa96}.

\subsection{Time-Dependent Schr\"{o}dinger Equation}

Direct numerical solution of the time-dependent Schr\"{o}dinger equation for
laser-induced processes has come to be called \textquotedblleft
TDSE\textquotedblright. Numerical methods have advanced to the point that
accurate TDSE calculations can be performed over a wide range of laboratory
parameters. TDSE is regarded as \textquotedblleft exact\textquotedblright, and
it is frequently applied to verify the accuracy of analytical approximations.
A disadvantage of TDSE is that it does not give clear physical insights as to
why particular types of behavior occur in physical systems, whereas analytical
approximations can give rise to instructive physical interpretations.

The perception of exactness of the TDSE approach has been carried too far.
Many investigators (see, for example, Ref.\cite{jkp}) fail to notice the
low-frequency limit of the dipole approximation, and make the assumption that
TDSE can comfortably be extended all the way to zero frequency. The range of
applicability of the TDSE numerical method is indicated by the Dipole Oasis
domain in Fig.\ref{fg4}. Numerical methods applied to frequencies lower than
the Dipole Oasis would require eschewing the dipole approximation for
frequencies somewhat less than the low frequency limit of the Tunnel Oasis,
and full solution of the Dirac equation in three spatial coordinates for still
lower frequencies. Those capabilities do not exist at present.

\subsection{Length gauge and velocity gauge}

The meaning of \textquotedblleft length gauge\textquotedblright\ is
unambiguous: it is identical to the GM gauge. That is, it refers to the
representation of an electromagnetic field of laser origin by the
$\mathbf{r\cdot E}\left(  t\right)  $ scalar potential of Eq.(\ref{a}). In
other words, it approximates a PW field by the conceptually simpler QSE field.

The term \textquotedblleft velocity gauge\textquotedblright\ can be
misinterpreted, as detailed in the above discussion about the SFA. The
simplest solution to this situation is to confine the meaning of velocity
gauge to that use of the dipole approximation that is exactly gauge-equivalent
to the GM (or length) gauge. The basic interaction Hamiltonian for the
velocity gauge is, in atomic units and for a single particle:%
\begin{equation}
H_{I}^{VG}=\mathbf{A}\left(  t\right)  \mathbf{\cdot p}+\frac{1}{2}%
\mathbf{A}^{2}\left(  t\right)  . \label{m}%
\end{equation}

As the dipole approximation is employed in the SFA of
Refs.\cite{hr80,hr90,hrrev,crawdiss}, a suitable nomenclature is to call it
\textquotedblleft radiation gauge\textquotedblright\ or \textquotedblleft
radiation gauge\ in the dipole approximation\textquotedblright\ when that
modifier is appropriate. The terminology \textquotedblleft velocity
gauge\textquotedblright\ can convey the wrong impression.

\section{Summary}

For a charged particle in a laser field (or in any propagating field), it is
at low frequencies that the magnetic component of the field becomes most
important. This stands in sharp contrast to charged particle behavior in any
field describable by a scalar potential of the form $\mathbf{r\cdot E}\left(
t\right)  $ (a longitudinal field) for which no magnetic field exists. The
tunneling model of ionization is confined entirely to longitudinal fields, so
it has no relevance for low-frequency laser fields.

Quality indices based on the $\mathbf{r\cdot E}\left(  t\right)  $ potential
are actually counter-indicative. A frequently-employed example is the concept
that an analytical method should reproduce static-electric-field properties as
the frequency declines.\textbf{ }This violates the very concept of a
propagating field that, to maintain its propagation property, must always
retain the appropriate time dependence given by the phase of the propagating field.

The low-frequency failure of dipole-approximation methods applies to all such
techniques employed for laser fields, including numerical methods for solution
of the Schr\"{o}dinger equation (TDSE) and all length-gauge and velocity-gauge
treatments. The tunneling model is a length-gauge approximation that has the
additional limitation imposed by the high-frequency constraint that the energy
of a single photon of the field must be much less than the binding potential
of a prospective detached electron. The result is that the \textquotedblleft
Tunnel Oasis\textquotedblright\ of Fig.\ref{fg4}, defining the domain of laser
parameters for which the tunneling model can be applied, is far smaller than
the overall \textquotedblleft Dipole Oasis\textquotedblright\ in Fig.\ref{fg4}
within which the dipole approximation for laser-field effects has validity.

The Oases shown in Fig.\ref{fg4} continue to expand as the intensity falls
below the $3.5\times10^{12}W/cm^{2}$ lowest intensity of the figure.
Traditional AMO physics is conducted at much lower intensities where the
low-frequency failures of the GM gauge and of the tunneling model are not in
evidence. This is the most likely explanation for why the low-frequency
failure of the dipole approximation has not previously attracted notice. That
situation is changing.

Existing laser experiments at low frequencies already show the failure of the
tunneling approximation.


\begin{thebibliography}{99}                                                                                               %


\bibitem {gamow}Gamow G 1928 \textit{Z. Physik} \textbf{51} 204-212

\bibitem {oppie}Oppenheimer J R 1928 \textit{Phys. Rev. \ }\textbf{31} 66-81

\bibitem {keldysh}Keldysh L V 1965 \textit{Sov. Phys.--JETP} \textbf{20} 1307-14

\bibitem {n+r}Nikishov A\ I and Ritus V\ I 1966 \textit{Sov. Phys.--JETP}
\textbf{23} 168-177

\bibitem {ppt}Perelomov A M, Popov V\ S and Terent'ev M\ V 1966 \textit{Sov.
Phys.--JETP} \textbf{23} 924-934

\bibitem {jackson}Jackson J D 1975 \textit{Classical Electrodynamics} 2nd edn
(New York: Wiley)

\bibitem {gm}G\"{o}ppert-Mayer M 1931 \textit{Ann. Phys., Lpz. }\textbf{9} 273-94

\bibitem {hr62b}Reiss H\ R 1962 \textit{J. Math. Phys. }\textbf{3} 387-395

\bibitem {hr89}Reiss H\ R 2014 \textit{Phys. Rev. }A \textbf{89} 022116

\bibitem {jkp}Joachain C J, Kylstra N J and Potvliege R M 2012 \textit{Atoms
in Intense Laser Fields} (Cambridge: Cambridge)

\bibitem {arissian}Arissian L, Smeenk C, Turner F, Trallero C, Sokolov A V,
Villeneuve D\ M, Staudte A and Corkum P B 2010 \textit{Phys. Rev. Lett.}
\textbf{105} 133002

\bibitem {schwinger}Schwinger J 1951 \textit{Phys. Rev. }\textbf{82} 664-79

\bibitem {s+s}Sarachik E S and Schappert G T 1970 \textit{Phys. Rev. }D
\textbf{1} 2738-53

\bibitem {hrjmo}Reiss H\ R 2012 \textit{J. Mod. Opt. }\textbf{59} 1371-1383;
2013 \textbf{60} 687

\bibitem {hrjpb13}Reiss H\ R 2013 \textit{J. Phys. B: At. Mol. Opt. Phys.
}\textbf{46 }175601

\bibitem {hr62a}Reiss H R 1962 \textit{J. Math. Phys. }\textbf{3} 59-67

\bibitem {sengupta}Sengupta N D 1952 \textit{Bull. Math. Soc. (Calcutta)
}\textbf{44} 175-80

\bibitem {n+r64}Nikishov A\ I and Ritus V\ I 1964 \textit{Sov. Phys.--JETP}
\textbf{19} 529

\bibitem {l+l}Landau L D and Lifshitz E\ M\ 1975 \textit{The Classical Theory
of Fields }(Oxford: Pergamon)

\bibitem {hrrel}Reiss H\ R 1990 \textit{J. Opt. Soc. Am. }B \textbf{7} 574-86

\bibitem {t+d}Titi A\ S and Drake G\ W\ F 2012 \textit{Phys. Rev. }A
\textbf{85} 041404(R)

\bibitem {hr87}Reiss H\ R 2013 \textit{Phys. Rev. }A \textbf{87} 033421

\bibitem {smeenk}Smeenk C\ T\ L, Arissian L, Zhou B, Mysyrowicz A, Villenueve
D\ M, Staudte A and Corkum P\ B 2011 \textit{Phys. Rev. Lett. }\textbf{106} 193002

\bibitem {hr82}Reiss H\ R 2010 \textit{Phys. Rev. }A \textbf{82} 023418

\bibitem {shakeshaft}Shakeshaft R, Potvliege R\ M, D\"{o}rr M and Cooke W\ E
1990 \textit{Phys. Rev. A }\textbf{42} 1656-1668

\bibitem {sanguine}http://en.wikipedia.org/wiki/Project\_Sanguine

\bibitem {kmp}Popov V\ S, Karnakov B\ M and Mur V\ D 2004
\textit{JETP\ Letters }\textbf{79} 262-7

\bibitem {laval}Xiong W, Yergeau F, Chin S\ L and Lavigne P 1988 \textit{J.
Phys. B: At. Mol. Opt. Phys. }\textbf{21} L159-64

\bibitem {hr102}Reiss H\ R 2009 \textit{Phys. Rev. Lett.} \textbf{102} 143003

\bibitem {osu1}Colosimo P, Duomy G, Blaga C\ I, Wheeler J, Haury C, \textit{et
al.} 2008 \textit{Nat. Phys. }\textbf{4} 386-9

\bibitem {osu2}Blaga C\ I, Catoire F, Colosimo P, Paulus G G, Muller H\ G,
Agostini P and DiMauro L\ F 2009 \textit{Nat. Phys. }\textbf{5} 335-8

\bibitem {cbb}Corkum P B, Burnett N H and Brunel F 1989 \textit{Phys. Rev.
Lett. }\textbf{62} 1259-1262

\bibitem {dk}Delone N\ B and Krainov V\ P 1991 \textit{J. Opt. Soc. Am. }B
\textbf{8} 1207-1211

\bibitem {hr80}Reiss H\ R 1980 \textit{Phys. Rev. }A \textbf{22} 1786-1813

\bibitem {hr90}Reiss H\ R 1990 \textit{Phys. Rev. }A \textbf{42} 1476-1486

\bibitem {bondar}Bondar D\ I, Spanner M, Liu W\ K and Yudin G\ L 2009
\textit{Physical Review }A \textbf{79} 063404

\bibitem {hrjosa96}Reiss H\ R 1996 \textit{J. Opt. Soc. Am. }B \textbf{13} 355-362

\bibitem {hrrev}Reiss H\ R 1992 \textit{Prog. Quantum Electron. }\textbf{16} 1-71

\bibitem {crawdiss}Crawford D\ P 1994 \textit{PhD Dissertation: Relativistic
Ionization with Intense Linearly Polarized Light} (Washington, DC: American University)
\end{thebibliography}
\end{document}